\documentclass[aps,pre,preprint,groupedaddress,showpacs]{revtex4}
\usepackage{graphicx}
\begin{document}
\newcommand{\ud}{{\mathrm d}}

\title{Gain in Stochastic Resonance:
Precise Numerics versus Linear Response Theory beyond the Two-Mode
Approximation}


\author{Jes\'us Casado-Pascual, Claus Denk, Jos\'e  G\'omez-Ord\'o\~nez,
Manuel Morillo}
\affiliation{F\'{\i}sica Te\'orica,
Universidad de Sevilla, Apartado de Correos 1065, Sevilla 41080, Spain}
\author{Peter H\"anggi}
\affiliation{Institut f\"ur Physik,
Universit\"at Augsburg - Universit\"atsstra\ss e 1, D-86135 Augsburg,
Germany}

\date{\today}

\begin{abstract}
In the context of the phenomenon of {\em Stochastic Resonance} (SR) we
study the correlation function, the signal-to-noise ratio (SNR) and the
ratio of output over input SNR, i.e. the {\em gain}, which is associated
to the nonlinear response of a bistable system driven by time-periodic
forces and white Gaussian noise. These quantifiers for SR are evaluated
using the techniques of Linear Response Theory (LRT) beyond the usually
employed two-mode approximation scheme. We analytically demonstrate
within such an extended LRT description that the gain can indeed {\em
not} exceed unity. We implement an efficient algorithm, based on work by
Greenside and Helfand (detailed in the Appendix), to integrate the driven
Langevin equation over a wide range of parameter values. The predictions
of LRT are carefully tested against the results obtained from numerical
solutions of the corresponding Langevin equation over a wide range of
parameter values. We further present an accurate procedure to evaluate
the distinct contributions of the coherent and incoherent parts of the
correlation function to the SNR and the gain. As a main result we show
for subthreshold driving that both, the correlation function and the
SNR can deviate substantially from the predictions of LRT and yet, the
gain can be either larger or smaller than unity. In particular, we find
that the gain can exceed unity in the strongly nonlinear regime which is
characterized by weak noise and very slow multifrequency {\em
subthreshold} input signals with a small duty cycle. This latter result
is in agreement with recent analogue simulation results by Gingl {\em et
al.} in Refs.  \cite{ginmak00,ginmak01}.

\end{abstract}


\maketitle
\section{Introduction}
Over the last twenty years or so, a large amount of work has been
devoted to the study of the dynamics of noisy nonlinear systems
driven by external periodic forces. One of the main reasons for
this interest is related to the phenomenon of Stochastic Resonance
(SR) \cite{PT96,RMP,Wiesenfeld98,Anish99,Chemphyschem02} namely,
the possibility of using the concerted action of noise and
nonlinearity to augment selectively, for some parameter values,
the output of the nonlinear system with respect to what it would
be for a linear system dynamics.

The two common quantifiers for Stochastic Resonance are the
spectral amplification measure \cite {RMP,EPL89,PRA91} and the
signal-to-noise ratio (SNR) \cite{RMP,McNWie89}. They are defined
in terms of the Fourier components of the correlation function
associated to the stochastic variable, $x(t)$. Due to the
periodicity of the driving force, the stochastic process $x(t)$ is
explicitly  nonstationary. Thus, the two-time function $\langle
x(t+\tau) x(t) \rangle$ depends on both $t$ and $\tau$. For very
large values of $t$, this quantity is periodic in $t$ with the
period of the external driving. Thus, its cycle average over one
period of $t$ yields a function of just $\tau$: the correlation
function, $C(\tau)$. The analysis of its structure reveals that
$C(\tau)$ is the sum of two terms \cite{RMP}: One term is periodic
in $\tau$ with the same period as the driving force and it is
called the coherent part, $C_{coh}(\tau)$. The other term, the
incoherent part $C_{incoh}(\tau)$, decays to zero for
$\tau\rightarrow \infty$. The SNR of the output process, $x(t)$,
is defined as the ratio of the amplitude of the Fourier mode of
the coherent part at the driving frequency, and the power spectral
density of the incoherent part taken also at the driving
frequency. By definition, the SNR is thus a dimensional quantity.

The SNR of an input signal (SNR$|_{inp}$) containing the sum of
the external driving and the Gaussian white noise can easily be
evaluated. A convenient dimensionless parameter can then be
defined as follows: the gain, $G$, defined by the ratio of SNR of
the output over the  (SNR$|_{inp}$). For the case that the
Langevin dynamics is linear in $x$ driven by additive white
Gaussian noise, then the output SNR is exactly the same as
SNR$|_{inp}$; i.e. the gain assumes precisely the value unity. In
a general nonlinear case, neither the SNR nor the gain can be
evaluated exactly by analytical means. As a consequence, their
evaluation necessarily requires approximate procedures.

It was pointed out previously that the gain of a noisy nonlinear
dynamical system subject to subthreshold signals can not exceed
$1$ \cite{dykluc95,dewbia95}. This feature has been rationalized
using the ideas of Linear Response Theory (LRT), thought to be
valid for weak driving amplitudes and not too small noise
strengths. It should be pointed out, however, that the validity of
LRT critically depends also on the value of the frequency, as has
 convincingly been demonstrated in recent works
\cite{EPLcasado,FNLcasado}.

In the context of LRT theory it has been pointed out in \cite{dykluc95}
that a corollary of LRT is that ``for small amplitude signals, the
signal-to-noise ratio at the output of a system driven by a stationary
Gaussian noise does not exceed that at the input, even if the system
displays SR''. Moreover, in Ref.  \cite{dewbia95}, the authors state
that for ``small signal in a Gaussian noise background, it is a theorem
that the SNR at the output of a non-linear device must be less than or
equal to the SNR at the input''. On the other hand, studies on {\em
nondynamical} systems \cite{Kiss96,Chap97,Kiss00,Liu01}, on dynamical
systems driven by large amplitude sinusoidal forces \cite{haninc00}, and
on dynamical systems driven by pulsed (multifrequency) periodic forces
with subthreshold amplitudes \cite{ginmak00,ginmak01}, have reported
gains larger than unity.  Clearly, for this to occur, the stochastic
system must operate in a regime where LRT does not apply. It is
therefore of interest to delineate carefully the limit of applicability
of the LRT description of the correlation function, the SNR and the gain
of a nonlinear noisy driven system.

In this paper we have tackled this challenge by carrying out a
detailed numerical evaluation of the correlation function
$C(\tau)$ and its components, $C_{coh}(\tau)$ and
$C_{incoh}(\tau)$, of the SNR and the gain of a bistable noisy
system which is driven by time periodic forces. The numerical
predictions have been compared with those provided by the LRT
approximation which accounts for the full spectrum of all
relaxation modes.

As it is well known, LRT requires the knowledge of the system
susceptibility, or alternatively, of the correlation function of
the noisy system in the absence of driving,
$K(t)$~\cite{RMP,PRHT82,EPLcasado,FNLcasado,junhan93,Morillo95}.
None of these quantities are known exactly for nonlinear systems.
For sufficiently small values of the noise strength, suitable
analytical approximations to $K(t)$ can been used
\cite{RMP,junhan93,EPLcasado,PRHT82,FNLcasado}. On the other hand, for
large values of the noise intensity, we have evaluated $K(t)$ from
the numerical solution of the Fokker-Planck equation using an
adaptation of the split operator technique of Feit {\em et al.}
\cite{feifle82}, as it has been  detailed in \cite{morgom92}. In
this paper, we also present a detailed proof of the statement that
within LRT, the gain G$_{LRT}\le 1$, by use of the full spectral
approach; this proof differs from alternative attempts in Refs.
\cite{dykluc95, dewbia95} which use additional restrictions such
as a linear response theory for the fluctuations themselves.

The ``typical'' procedure to evaluate the SNR involves the Fourier
analysis of a very long record of the stochastic trajectory,
$x(t)$. Using the Fast Fourier Transform (FFT) of the record, the
corresponding periodogram is constructed. There are several
drawbacks with this procedure. There are subtleties inherent to
the interpretation and evaluation of the periodogram (see for
instance the critical comments in \cite{numrec}). There are also
major problems associated with the fact that the power spectrum
contains $\delta$-peaks at the driving frequency and its higher
harmonics arising from the coherent part of the correlation
function.  The contribution of the incoherent part at those
frequencies is embedded in those peaks, and it is not a simple
task to estimate the separate contribution to the peaks of the
coherent and incoherent parts of the periodogram. The evaluation
of the SNR-gain requires a good knowledge of both contributions,
and any small error in the estimation of the incoherent
contribution, yields unreasonable values for the gain. Indeed, in
our opinion, a much better estimate would be obtained if the
periodic part of the output signal were subtracted from the data
before performing its FFT.

In this work we propose such an alternative procedure. The
Langevin equation is numerically integrated for a large number of
noise realizations. The time evolution of the correlation function
and its coherent part are directly evaluated from the numerical
solution after averaging over the noise realizations. The
incoherent part is obtained from the difference
$C_{incoh}(\tau)=C(\tau)-C_{coh}(\tau)$. As the definition of SNR
requires just the amplitude of the Fourier mode of $C_{coh}(\tau)$
and the spectral density of $C_{incoh}(\tau)$ at the driving
frequency, the SNR can  readily  be evaluated with two numerical
quadratures; i.e. there is no need to construct the full spectrum.

The paper is organized as follows. In the next Section we introduce the
model and provide definitions of the quantities of interest. In Section
III, the main points of the LRT description of the correlation functions
are detailed. We also present in this Section a novel and
straightforward proof of the fact that G$_{LRT}\le 1$, based on the
spectral properties of the Fokker-Planck operator, and its adjoint, in
the absence of driving. In Section IV, we present the numerical
procedure used to obtain the correlation function, the SNR and the gain
from the numerical solution of the Langevin equation. The very efficient
algorithm used in this work is summarized in the Appendix. The numerical
results are compared with the predictions of LRT for a variety of
parameters and two distinct types of driving forces: a monochromatic
force and a periodic sequence of pulses. Finally, we present conclusions
for the main findings of our work.

\section{Correlation function,  signal-to-noise ratio and  gain}
Let us consider a system characterized by a single degree of freedom,
$x$, subject to the action of a zero average Gaussian white noise with
$\langle \xi(t)\xi(s)\rangle = 2D\delta(t-s)$ and driven by an
external periodic signal $F(t)$ with period $T$.  In the Langevin
description, its dynamics is generated by the equation
\begin{equation}
\label{langev} \dot{x}(t)=-U'\left[ x(t) \right]+F(t)+\xi(t).
\end{equation}
The corresponding linear Fokker-Planck equation (FPE) for the
probability density $P(x,t)$ reads
\begin{equation}
\label{FP}
\frac{\partial}{\partial t}P(x,t)={\hat {\mathcal L}}(t) P(x,t),
\end{equation}
where
\begin{equation}
{\hat {\mathcal L}}(t)=\frac{\partial}{\partial x}\left[ U'(x)-F(t)+ D
\frac{\partial}{\partial x}\right].
\end{equation}
In the expressions above, $U'(x)$ represents the derivative of the
potential $U(x)$. The periodicity of the external driving $F(t)$
allows its Fourier series expansion in the harmonics of the
fundamental frequency $\Omega=2\pi/T$, i.e.,
\begin{equation}
\label{fourf}
F(t)=\sum_{n=1}^\infty \left [ f_n \cos (n \Omega t) + g_n \sin
(n \Omega t) \right ],
\end{equation}
with the Fourier coefficients, $f_n$ and $g_n$, given by
\begin{eqnarray}
f_n&=&\frac 2T \, \int_0^T \ud t\, F(t) \cos (n \Omega t)
\nonumber \\
g_n&=&\frac 2T \, \int_0^T \ud t\, F(t) \sin (n \Omega t).
\end{eqnarray}
Here, we are assuming that the cycle average of the external
driving over its period equals zero.

The two-time correlation function $\langle x(t+\tau)
x(t)\rangle_{\infty}$ in the limit $t \rightarrow \infty$ is given
by
\begin{equation}
\langle x(t+\tau) x(t)\rangle_{\infty} =\int_{-\infty}^{\infty} \ud x'
\,x' P_{\infty}(x',t) \int_{-\infty}^{\infty}
\ud x \,x P_{1|1}(x,t+\tau|x',t),
\end{equation}
where $P_{\infty}(x,t)$ is the time-periodic, asymptotic long time
solution of the FPE and the quantity $P_{1|1}(x,t+\tau|x',t)$
denotes the two-time conditional probability density that the
stochastic variable will have a value near $x$ at time $t+\tau$ if
its value at time $t$ was exactly $x'$. It can been shown
\cite{RMP,PRA91} that, in the limit $t \rightarrow \infty$, the
two-time correlation function $\langle x(t+\tau)
x(t)\rangle_\infty$ becomes a periodic function of $t$ with the
period of the external driving. Then, we define the one-time
correlation function, $C(\tau)$, as the average of the two-time
correlation function over a period of the external driving, i.e.,
\begin{equation}
\label{ctau}
C(\tau)= \frac{1}{T} \int_{0}^{T} \ud t \, \langle x(t+\tau)
x(t)\rangle_{\infty}.
\end{equation}
The correlation function $C(\tau)$ can be written exactly as the
sum of two contributions: a coherent part, $C_{coh}(\tau)$, which
is periodic in $\tau$ with period $T$, and an incoherent part
which decays to $0$ for large $\tau$. The coherent part
$C_{coh}(\tau)$ is given by \cite{RMP,PRA91}
\begin{equation}
\label{chtau}
C_{coh}(\tau) =\frac{1}{T} \int_{0}^{T} \ud t \,\langle x(t+\tau)
\rangle_{\infty} \langle x(t) \rangle_{\infty},
\end{equation}
where $\langle x(t) \rangle_{\infty}$ is the average value evaluated
with the asymptotic form of the probability density $P_{\infty}(x,t)$.

It is possible to carry out a formal analysis of $C(\tau)$ and its
coherent and incoherent components by making use of the spectral
analysis of the Floquet operator associated with the Fokker-Planck
dynamics. But an explicit evaluation of the correlation function
is generally impossible; thus,  one has to rely on numerical
results obtained from integrating either the Langevin or the FPE,
or by use of  approximate analytical descriptions.

According to McNamara and Wiesenfeld \cite{McNWie89}, the SNR is
defined in terms of the Fourier transform of the coherent and
incoherent parts of $C(\tau)$. As the correlation function is even
in time and we evaluate its time dependence for $\tau \ge 0$, it
is convenient to use its Fourier cosine transform, defined as
\begin{equation}
\label{fourier}
\tilde{C}(\omega)=\frac 2\pi \int_0^\infty \ud\tau\,C(\tau) \cos (\omega
\tau);\; C(\tau)=\int_0^\infty \ud\omega\, \tilde{C}(\omega) \cos (\omega
\tau).
\end{equation}
For the SNR we define:
\begin{equation}
\label{snr}
SNR =\frac {\lim_{\epsilon \rightarrow 0^+}
\int_{\Omega-\epsilon}^{\Omega+\epsilon} \ud\omega\;
\tilde{C}(\omega)}{\tilde{C}_{incoh}(\Omega)}.
\end{equation}
Note that this definition of the SNR differs by a  factor $2$,
stemming from the same contribution at $\omega = - \Omega$, from
the definitions used in earlier works \cite {RMP,PRA91}. The
periodicity of the coherent part gives rise to delta peaks in the
spectrum. Thus, the only contribution to the numerator in Eq.\
(\ref{snr}) stems from the coherent part of the correlation
function. The evaluation of the SNR requires the knowledge of the
Fourier components of $C_{coh}(\tau)$ and $C_{incoh}(\tau)$ at the
fundamental frequency of the driving force. The entire Fourier
spectrum is not needed. The evaluation of the SNR thus requires
the evaluation of two well defined numerical quadratures only;
i.e.,
\begin{equation}
\label{snr1}
SNR=\frac {{\frac 2T} \int_0^T \ud \tau \,C_{coh}(\tau) \cos (\Omega
\tau)}{ \frac 2\pi \int_o^\infty \ud \tau \,C_{incoh}(\tau) \cos (\Omega
\tau)}.
\end{equation}
The signal-to-noise ratio for an input signal $F(t)+\xi(t)$, is given by
\begin{equation}
\label{snrinp}
SNR|_{inp}=\frac{ \pi(f_1^2+g_1^2)}{4D}.
\end{equation}
The so-called gain is defined as the ratio of the SNR of the
output over the SNR of the input; namely,
\begin{equation}
\label{gain}
G=\frac {SNR}{SNR|_{inp}}.
\end{equation}

\section{Linear Response Theory  beyond the two-mode Approximation }

The Linear Response Theory provides a general procedure to
describe the correlation function in an approximate way.  The
basic quantity of LRT is the system response function, $\chi(t)$.
It is related to the equilibrium time correlation function of the
system in the absence of external driving, $K(t)$, via the
fluctuation-dissipation theorem (FDT)
\cite{RMP,PRA91,PRHT82,junhan93}, i.e.,
\begin{equation}
\label{fdt}
\chi(t)= \left \{ \begin{array}
{r@{\quad:\quad}l}
0 & t\le 0\\
-\frac 1D \dot{K}(t) & t > 0

\end{array}
\right .
\end{equation}
The equilibrium time correlation function $K(t)$ is defined as
\begin{equation}
\label{eqcorr}
K(t)= \int_{-\infty}^\infty \ud x'\, x' P^{(eq)}(x')\,\int_{-\infty}^\infty
\ud x\, xP^{(0)}_{1|1}(x,t|x'),
\end{equation}
where $P^{(eq)}(x)$ is the equilibrium distribution of the
non-driven system,
\begin{equation}
P^{(eq)}(x)=N e^{-\frac {U(x)}D },
\end{equation}
and $P^{(0)}_{1|1}(x,t|x')$ is the conditional probability density to
find, in the absence of driving, the variable near $x$ at time $t$, if
it was initially at exactly $x'$. Here we are assuming that the
potential $U(x)$ is even in $x$, so that $\langle x \rangle_{eq}=0$.

Within LRT, the long time average value $\langle
x(t)\rangle^{(LRT)}_\infty$ is given by
\begin{equation}
\label{lrt}
\langle x(t)\rangle^{(LRT)}_\infty =\int_0^\infty \ud\tau\, \chi(\tau)
F(t-\tau).
\end{equation}
Insertion of the Fourier expansion Eq.\ (\ref{fourf}) into Eq.\
(\ref{lrt}) leads to
\begin{equation}
\label{lrt1}
\langle x(t)\rangle^{(LRT)}_\infty = \sum_{n=1}^\infty  \left [
M_n^{(LRT)} \cos (n \Omega t)  + N_n^{(LRT)}  \sin (n \Omega t) \right ],
\end{equation}
where the coefficients $M_n^{(LRT)}$ and $N_n^{(LRT)}$ are given by
\begin{equation}
\label{coeflrt}
M_n^{(LRT)} = f_n \chi_n^{(r)} -g_n \chi_n^{(i)}; \quad N_n^{(LRT)} =
f_n \chi_n^{(i)} + g_n \chi_n^{(r)}.
\end{equation}
In these formulas, we have introduced the quantities $\chi_n^{(r)}$ and
$\chi_n^{(i)}$ defined as
\begin{eqnarray}
\label{suscepa}
\chi_n^{(r)} &=& \int_0^\infty \ud\tau\, \chi(\tau) \cos (n \Omega \tau)\\
\label{suscepb}
\chi_n^{(i)} &=& \int_0^\infty \ud\tau\, \chi(\tau) \sin (n \Omega \tau).
\end{eqnarray}
The use of the FDT in the above expressions allows us to write immediately
\begin{eqnarray}
\label{suscep1}
\chi_n^{(r)} &=& \frac {\langle x^2 \rangle_{eq}- n\Omega \,
\int_0^\infty \ud t\, K(t) \sin (n \Omega t)}D \\
\label{suscep2}
\chi_n^{(i)} &=& \frac {n \Omega}D \, \int_0^\infty \ud t\, K(t) \cos (n
\Omega t).
\end{eqnarray}
It then follows from Eq.\ (\ref{chtau}) that within LRT, the coherent
part of the correlation function is given by,
\begin{equation}
\label{cohlrt}
C_{coh}^{(LRT)}(\tau)=\frac 12\sum_{n=1}^\infty \left [
\left(M_n^{(LRT)}\right)^2 +\left(N_n^{(LRT)}\right)^2 \right ] \cos( n
\Omega \tau).
\end{equation}
As discussed in Refs.\cite{RMP,PRA91,PRHT82}, LRT amounts to keeping the
leading term in the perturbation treatment of the dynamics of the
stochastic process $x(t)$ in powers of the driving amplitude.
Then, within the spirit of perturbation theory, the leading term
in the expansion of the incoherent part corresponds to the
correlation function of the system in the absence of driving
force, i.e., $C_{incoh}^{(LRT)}(\tau)=K(\tau)$.

Taking into account that $C_{coh}^{LRT}(\tau)$ is periodic in $\tau$, it
follows from Eqs.~(\ref{fourier}) and (\ref{cohlrt}) that
\begin{equation}
\label{cohlrtom}
\tilde{C}_{coh}^{(LRT)}(\omega)= \frac 12\sum_{n=1}^\infty \left [
\left(M_n^{(LRT)}\right)^2 +\left(N_n^{(LRT)}\right)^2 \right ] \left [
\delta (n\Omega -\omega)+\delta (n\Omega +\omega) \right ].
\end{equation}
Thus, it follows from the definition of the SNR, Eq.~(\ref{snr}),
that, within LRT, we have
\begin{eqnarray}
\label{snrlrt}
SNR^{(LRT)}&=& \frac { \frac 12 \left [ \left(M_1^{(LRT)}\right)^2
+\left(N_1^{(LRT)}\right)^2 \right ]}{\tilde {K}(\Omega)}\nonumber \\
&=&\frac {\pi \Omega \left(f_1^2+g_1^2
\right)\left[\left(\chi_1^{(r)}\right)^2+\left(\chi_1^{(i)}
\right)^2\right]}{4 D\chi_1^{(i)}},
\end{eqnarray}
where $\tilde {K}(\Omega)$ is the Fourier cosine transform of
$K(t)$, defined according to Eq.~(\ref{fourier}). In arriving at
Eq.~(\ref{snrlrt}) we  have also used Eqs.~(\ref{coeflrt}),
(\ref{suscepa}), (\ref{suscepb}) and (\ref{suscep2}).

Taking into account Eqs.~(\ref{snrinp}), (\ref{gain}) and
(\ref{snrlrt}), one readily finds  that the gain within LRT is
given by
\begin{equation}
\label{gainlrt}
G^{(LRT)}= \frac {SNR^{(LRT)}}{SNR |_{inp}}= \frac { \Omega
\left[\left(\chi_1^{(r)}\right)^2
+\left(\chi_1^{(i)}\right)^2\right]}{\chi_1^{(i)}}.
\end{equation}
This is a general expression for $G^{(LRT)}$ valid for any shape
of periodic driving.

The last expression will allow us to show  that $G^{(LRT)}$ can,
indeed, not exceed unity. Although this assertion has been
discussed previously in Ref.~\cite{dykluc95,dewbia95}, we next
will present a detailed and hopefully very clear proof for this
prominent assertion.

As shown in the Appendix of Ref.~\cite{RMP}, see also in Ref.
\cite{EPLcasado,FNLcasado}, the susceptibility, $\chi(t)$, can be
expressed as
\begin{equation}
\chi(t) = - \sum_{p=1}^\infty e^{-\lambda_p t} \langle 0|x|p\rangle
\langle p | \frac {\partial }{\partial x} | 0\rangle,
\end{equation}
where $|p\rangle = \psi_p(x)$, $\langle p |=\psi_p^{\dagger}(x)$ and
$\lambda_p$ are the eigenfunctions and eigenvalues of the FP operator
${\hat \mathcal{L}}_0$ associated to the undriven dynamics and its adjoint,
${\hat \mathcal{L}}_0^{\dagger}$, i.e.,
\begin{equation}
\label{eig}
{\hat \mathcal{L}}_0 \psi_p(x)=-\lambda_p \psi_p(x), \quad {\hat
\mathcal{L}}_0^{\dagger} \psi_p^{\dagger}(x)=-\lambda_p
\psi_p^\dagger(x).
\end{equation}
Using the above representation of the susceptibility in Eqs.~(\ref{suscepa})
and (\ref{suscepb}) with $n=1$, we find
\begin{eqnarray}
\label{uno}
\chi_1^{(r)} &=& -\sum_{p=1}^\infty \frac {\lambda_p}{\lambda_p^2 +
\Omega^2}\, \langle 0|x|p\rangle \langle p |\frac {\partial }{\partial
x} | 0\rangle=\sum_{p=1}^\infty \frac {\lambda_p}{\lambda_p^2 +
\Omega^2}\, \left|\langle 0|x|p\rangle \langle p |\frac {\partial
}{\partial x} | 0\rangle\right|, \\
\label{dos}
\chi_1^{(i)} &=& -\sum_{p=1}^\infty \frac {\Omega}{\lambda_p^2 +
\Omega^2}\, \langle 0|x|p\rangle \langle p |\frac {\partial }{\partial
x} | 0\rangle=\sum_{p=1}^\infty \frac {\Omega}{\lambda_p^2 + \Omega^2}\,
\left|\langle 0|x|p\rangle \langle p |\frac {\partial }{\partial x} |
0\rangle\right|.
\end{eqnarray}
Here, we have used the inequality
\begin{equation}
\label{ineq}
\langle 0|x|p\rangle \langle p |\frac {\partial }{\partial x} | 0\rangle
\le 0,
\end{equation}
which can be proved as
follows. Multiplying the first equation in Eq.\ (\ref{eig}) by $x$ and
carrying out an integration by parts, one obtains
\begin{eqnarray}
-\lambda_p \langle 0|x|p\rangle &=& \int_{-\infty}^{\infty} \ud x \,x\,
{\hat \mathcal{L}}_0 \psi_p(x)=-\int_{-\infty}^{\infty} \ud x \,U'(x)
\psi_0(x) \psi_p^{\dagger} (x)\nonumber\\ &=&D\int_{-\infty}^{\infty}
\ud x \, \psi_p^{\dagger}(x)\frac{\partial}{\partial x}
\psi_0(x)=D\,\langle p |\frac {\partial }{\partial x} |0\rangle,
\end{eqnarray}
where we have taken into account that $\psi_0(x)=P^{(eq)}(x)$ and
$\psi_p(x)=\psi_0(x)\psi_p^{\dagger}(x)$, so that
$\psi_0^{\dagger}(x)=1$. Therefore, $\langle 0|x|p\rangle \langle p
|\frac {\partial }{\partial x} | 0\rangle =-\lambda_p \left(\langle
0|x|p\rangle\right)^2/D \leq 0$.
Using in Eqs.~(\ref{uno}) and (\ref{dos}) the Cauchy-Schwarz inequality,
we find
\begin{eqnarray}
\label{tres}
\left(\chi_1^{(r)}\right)^2 &=&\left[\sum_{p=1}^\infty \frac {\lambda_p
\left|\langle 0|x|p\rangle \langle p |\frac {\partial }{\partial x} |
0\rangle\right|^{1/2}}{\lambda_p^2 + \Omega^2}\, \left|\langle
0|x|p\rangle \langle p |\frac {\partial }{\partial x} |
0\rangle\right|^{1/2}\right]^2\nonumber\\ &\leq& \sum_{p=1}^\infty\frac
{\lambda_p^2\left|\langle 0|x|p\rangle \langle p |\frac {\partial
}{\partial x} | 0\rangle\right| }{\left(\lambda_p^2 +
\Omega^2\right)^2}\,\sum_{q=1}^\infty \left|\langle 0|x|q\rangle\langle
q |\frac {\partial }{\partial x} | 0\rangle\right|,\\
\label{cuatro}
\left(\chi_1^{(i)}\right)^2 &=&\left[\sum_{p=1}^\infty \frac
{\Omega \left|\langle 0|x|p\rangle \langle
p |\frac {\partial }{\partial x} | 0\rangle\right|^{1/2}}{\lambda_p^2 + \Omega^2}\, \left|\langle
0|x|p\rangle \langle p |\frac {\partial }{\partial x} |
0\rangle\right|^{1/2}\right]^2\nonumber\\ &\leq& \sum_{p=1}^\infty\frac
{\Omega^2\left|\langle
 0|x|p\rangle \langle p |\frac {\partial }{\partial x} |
0\rangle\right|}{\left(\lambda_p^2 + \Omega^2\right)^2}\,\sum_{q=1}^\infty \left|\langle 0|x|q\rangle\langle q
|\frac {\partial }{\partial x} | 0\rangle\right|.
\end{eqnarray}
Taking into account that $\langle 0|x|0\rangle=0$ the completeness
relation yields $\sum_{q=1}^\infty \left|\langle 0|x|q\rangle\langle q
|\frac {\partial }{\partial x} | 0\rangle\right|=-\sum_{q=0}^\infty
\langle 0|x|q\rangle\langle q |\frac {\partial }{\partial x} |
0\rangle=-\langle 0|x\frac{\partial }{\partial x} | 0\rangle=1$.  Thus,
by adding Eq.~(\ref{tres}) to Eq.~(\ref{cuatro}), one obtains
\begin{equation}
\label{cinco}
\left(\chi_1^{(r)}\right)^2+\left(\chi_1^{(i)}\right)^2\leq\sum_{p=1}^\infty
\frac {\left|\langle 0|x|p\rangle \langle p |\frac {\partial }{\partial
x} | 0\rangle\right|}{\lambda_p^2 + \Omega^2}=\frac{\chi_1^{(i)}}{\Omega}.
\end{equation}
Finally, inserting Eq.~(\ref{cinco}) into (\ref{gainlrt}), we
obtain the seminal inequality that $G^{(LRT)}\leq 1$.

Put differently, the gain of a nonlinear system operating in a
regime where LRT provides a valid description cannot reach values
greater than $1$. This result is valid for {\em any} periodic
external driving. Notice that this finding does not preclude the
possibility of obtaining values for the SNR-gain larger than unity
when the conditions are such that the use of LRT is not sensible.

\section{Numerical results}

In this Section, we will carry out the numerical evaluation of the
different magnitudes defined above. Our goal is to compare the
predictions of LRT with the results obtained from the numerical solution
of the Langevin equation, Eq.\ (\ref{langev}). We will consider the dynamics
in the bistable potential $U(x)=-x^2/2+x^4/4$ driven by time periodic
forces.

The evaluation of the different magnitudes using LRT requires the
knowledge of $K(t)$ [cf.  Eqs.\ (\ref{coeflrt}, \ref{cohlrt},
\ref{snrlrt}, \ref{gainlrt})]. For nonlinear problems, explicit
expressions for $K(t)$ are unknown, but useful approximations have
been presented in the literature. For the bistable potential,
$U(x)=-x^2/2+x^4/4$, Jung and H\"anggi \cite{junhan93} have used
the two-mode approximation. It is based on the existence of a
large difference in the time scales associated to inter-well and
intra-well motions, and it is expected to be valid for small
values of the noise strength $D$. With this model, one finds
\begin{equation}
\label{corr2mode}
K(\tau)= g_1 e^{-\lambda_1 \tau}+g_2 e^{-\alpha \tau}
\end{equation}
where \cite{RMP}
\begin{equation}
\label{lambda1}
\lambda_1 \approx \frac {\sqrt 2}\pi\, (1-\frac 32 D)\,\exp(-1/(4D)),
\end{equation}
and $\alpha=2$. The weights, $g_1$ and $g_2$, can be obtained from the
moments of the equilibrium distribution in the absence of driving using the
expressions
\begin{equation}
\label{weight2}
g_2= \frac {\lambda_1 \langle x^2\rangle_{eq}}{\lambda_1 -\alpha} +
\frac { \langle x^2\rangle_{eq} - \langle x^4\rangle_{eq}}{\lambda_1 -\alpha}
\end{equation}
\begin{equation}
\label{weight1}
g_1=\langle x^2\rangle_{eq} -g_2.
\end{equation}
To leading order in $D$, we can replace $\lambda_1$ by
$\lambda_K=\sqrt 2/\pi \exp(-1/(4D))$, $g_1 \approx 1$ and $g_2
\approx D/ \alpha$. This is the limit considered in Ref.
\cite{gang92}. In the results reported below, we have also
considered values of $D$ so large that the two-mode approximation
becomes inadequate. Therefore, the correlation function in
the absence of driving has been evaluated numerically from the FPE
in the absence of driving following the procedure discussed in
Ref. \cite{morgom92}.

The numerical evaluation of the correlation function $C(\tau)$ and
its coherent and incoherent parts proceeds as follows. Stochastic
trajectories, $x^{(j)}(t)$, are generated by numerically
integrating the Langevin equation for every realization $j$
of the white noise $\xi(t)$, starting from a given initial
condition $x_0$.  The numerical solution is based on the algorithm
developed by Greenside and Helfand \cite{hel79,grehel81}. The
essence of the algorithm is briefly sketched in the Appendix.
After allowing for a relaxation transient stage, we start
recording the time evolution of each random trajectory for many
different trajectories. Then, we construct the two-time ($t$ and
$\tau$) correlation function, i.e.,
\begin{equation}
\langle x(t+\tau) x(t) \rangle_\infty = \frac 1N \sum_{j=1}^N
x^{(j)}(t+\tau) x^{(j)}(t),
\end{equation}
as well as the product of the averages
\begin{equation}
\langle x(t+\tau)\rangle_\infty \langle x(t) \rangle_\infty = \left (\frac 1N
\sum_{j=1}^N  x^{(j)}(t+\tau)  \right ) \left (\frac 1N
\sum_{j=1}^N x^{(j)}(t)  \right ),
\end{equation}
where $N$ is the number of stochastic trajectories considered. The
correlation function $C(\tau)$ and its coherent part
$C_{coh}(\tau)$ are then obtained using their definitions in
Eqs.~(\ref{ctau}) and (\ref{chtau}), performing the cycle average
over one period of $t$.  The difference between the values of
$C(\tau)$ and $C_{coh}(\tau)$ allows us to obtain the values for
$C_{incoh}(\tau)$. It is then straightforward to evaluate the
Fourier component of $C_{coh}(\tau)$ and the Fourier transform of
$C_{incoh}(\tau)$ at the driving frequency by numerical
quadrature. With that information, the numerator and the
denominator in the expression (\ref{snr}) for the SNR and the gain
(Eq.\ (\ref{gain})) are obtained.

We shall analyze two different types of  periodic driving forces.
First, let us consider the well known situation with a
monochromatic, single frequency force, $A\cos (\Omega t)$, with
amplitude strength $A$ and angular frequency $\Omega$ \cite{RMP}.
In this case, the formulas in Sect. III simplify considerably
because $f_1=A$, while all the other Fourier components of the
driving force vanish. The second case corresponds to a periodic
force with period $T$, with a sequence of pulses of length
$t_c<T/2$ ; namely,
\begin{equation}
\label{pulse}
F(t)= \left \{ \begin{array}
{r@{\quad:\quad}l}
A & 0\le t < t_c \\
-A& \frac T2 \le t < \frac T2+t_c\\
0& {\rm otherwise}.
\end{array}
\right .
\end{equation}
In this case, we have
\begin{equation}
f_1=\frac {2A}\pi\sin(\Omega t_c);  \quad g_1=\frac {2A}\pi \left [1-\cos
(\Omega t_c) \right ],
\end{equation}
where $\Omega=2\pi/T$ is the fundamental frequency. This force is
characterized by its amplitude, its period and its duty cycle, which is
defined as $2t_c/T$. Recently, Gingl {\em et al.}
\cite{ginmak00,ginmak01} have carried out analogue simulations of
systems that are subjected to wideband Gaussian noise and driving forces
of this second type. They report values for the gain which greatly
exceeds unity, for driving amplitudes below its threshold value. If this
is the case, then strong deviations from the LRT should be observed as
well.

\subsection{Monochromatic driving}

\begin{figure}
\includegraphics[width=10cm]{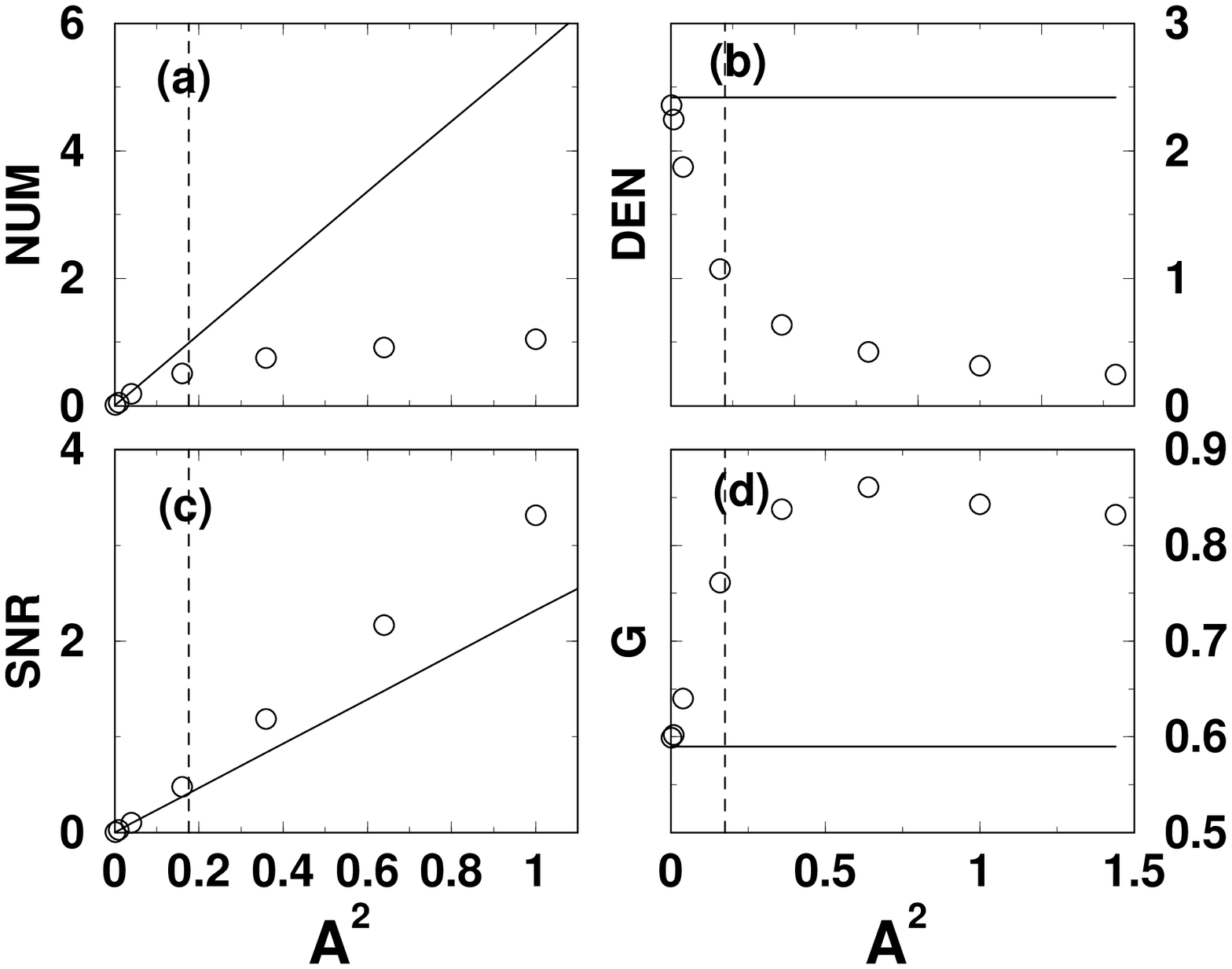}
\caption{\label{wp1}The dependence of several SR quantifiers versus the
square of the driving amplitude, $A^2$, given by LRT (solid line) and by
the numerical solution of the Langevin equation (circles). In panels (a)
and (b), we plot, respectively, the numerator and denominator appearing
in the definition of $SNR$, cf. Eq.\ (\ref{snr1}). The behaviors of the
$SNR$ and the gain are depicted in panels (c) and (d). The driving force
is monochromatic with frequency $\Omega=0.1$ and the white noise
strength is kept constant at the value $D=0.2$. In all panels, the
vertical dashed line indicates the square of the value of the dynamical
threshold amplitude, $A_{th}$, at the angular driving frequency,
$\Omega$. In panel (d), a dotted horizontal line is drawn at the gain
value of 1 as a guide to the eye.}
\end{figure}

In Fig.\ \ref{wp1} we depict the results obtained for a monochromatic
driving force with angular frequency $\Omega=0.1$, noise strength
$D=0.2$ and several values of the amplitude. In the deterministic
dynamics ($D=0$), an external periodic force with the indicated
frequency induces sustained oscillations between the minima of the
potential for $A \geq A_{th} \simeq 0.419$. Note that this nonadiabatic
frequency raises the threshold value for superthreshold driving beyond
its adiabatic lower limit of $A_{th}^{(ad)} = \sqrt{4/27} \simeq
0.3849$. Thus, we will take this value as the amplitude threshold value
at the frequency $\Omega=0.1$. In panel (a), we plot the {\em
numerators} of the SNR in Eq.\ (\ref{snr}) and of SNR$^{(LRT)}$ in Eq.\
(\ref{snrlrt}) {\em vs.} $A^2$. The solid straight line represents the
LRT result, while the circles correspond to the numerical results. The
graph reveals that for amplitude strengths $A < 0.1$ the predictions of
LRT match well the numerical results, as can be expected. When the
amplitude increases, the deviations of LRT from the precise numerical
results are large. LRT predicts a much larger amplification of the
output amplitude than the one obtained numerically.  In panel (b), we
plot the {\em denominators} of the SNR in Eq.\ (\ref{snr}) (circles) and
of SNR$^{(LRT)}$ in Eq.\ (\ref{snrlrt}) (solid line) {\em vs.} $A^2$.  In
LRT, the denominator is independent of $A$. Once again, the predictions
of LRT match the numerical results for $A < 0.1$. For larger values of
$A$, the influence of the driving amplitude on the relaxation of
$C_{incoh}(\tau)$ is very strong and the numerical results for the
denominator are much smaller than the ones obtained within LRT. It is
then clear that LRT will yield a valid description of the
signal-to-noise ratio for small driving amplitudes only as depicted in
panel (c). We notice that the values for the SNR provided by the
numerics are larger than SNR$^{(LRT)}$. This is so although linear
response theory predicts larger spectral amplifications, see in
\cite{PRA91}, of the average output than what really occurs. The
modifications in the behavior of the incoherent part of the correlation
function with respect to its behavior in the absence of driving are more
than enough to compensate for the behavior of the numerators. In panel
(d) we plot the gain {\em vs.} $A^2$. There exists
an optimum value for the driver amplitude ($A\sim 0.8$) at which the
gain becomes maximized. Nonetheless, the gain is always smaller than
unity. LRT requires that $G^{(LRT)}\le 1$.  These strong deviations of
the predictions of LRT about the behavior of the two components of the
correlation function with respect to the numerical results tell us that
LRT cannot be invoked to explain the fact that the gain is smaller than
$1$ for the range of parameter values considered in this figure; i.e. a
gain below $1$ occurs here within the nonlinear regime.

\begin{figure}
\includegraphics[width=10cm]{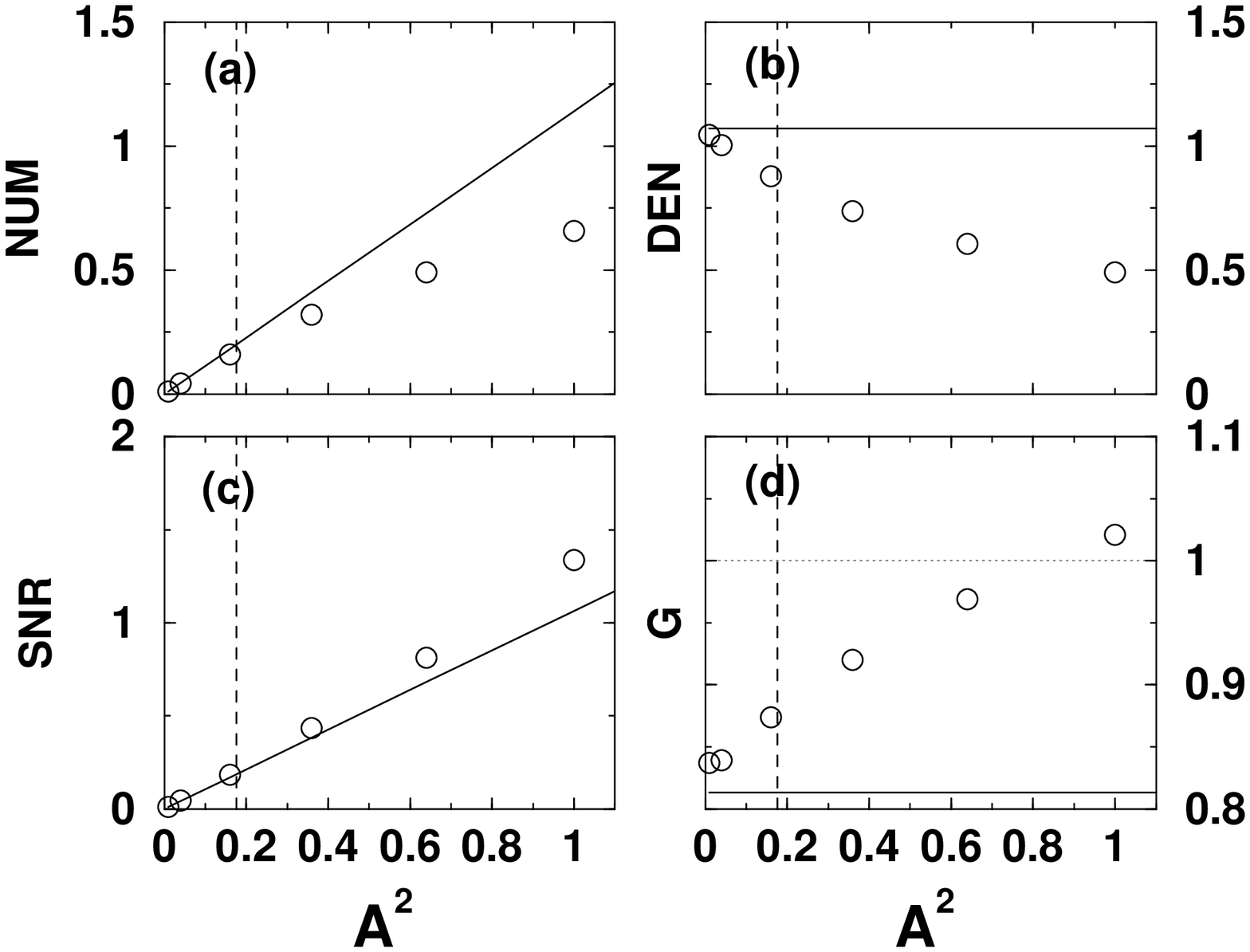}
\caption{\label{wp2} The same as in Fig.\ \ref{wp1} but now for
$\Omega=0.1$ and  $D=0.6$.}
\end{figure}

\begin{figure}
\includegraphics[width=10cm]{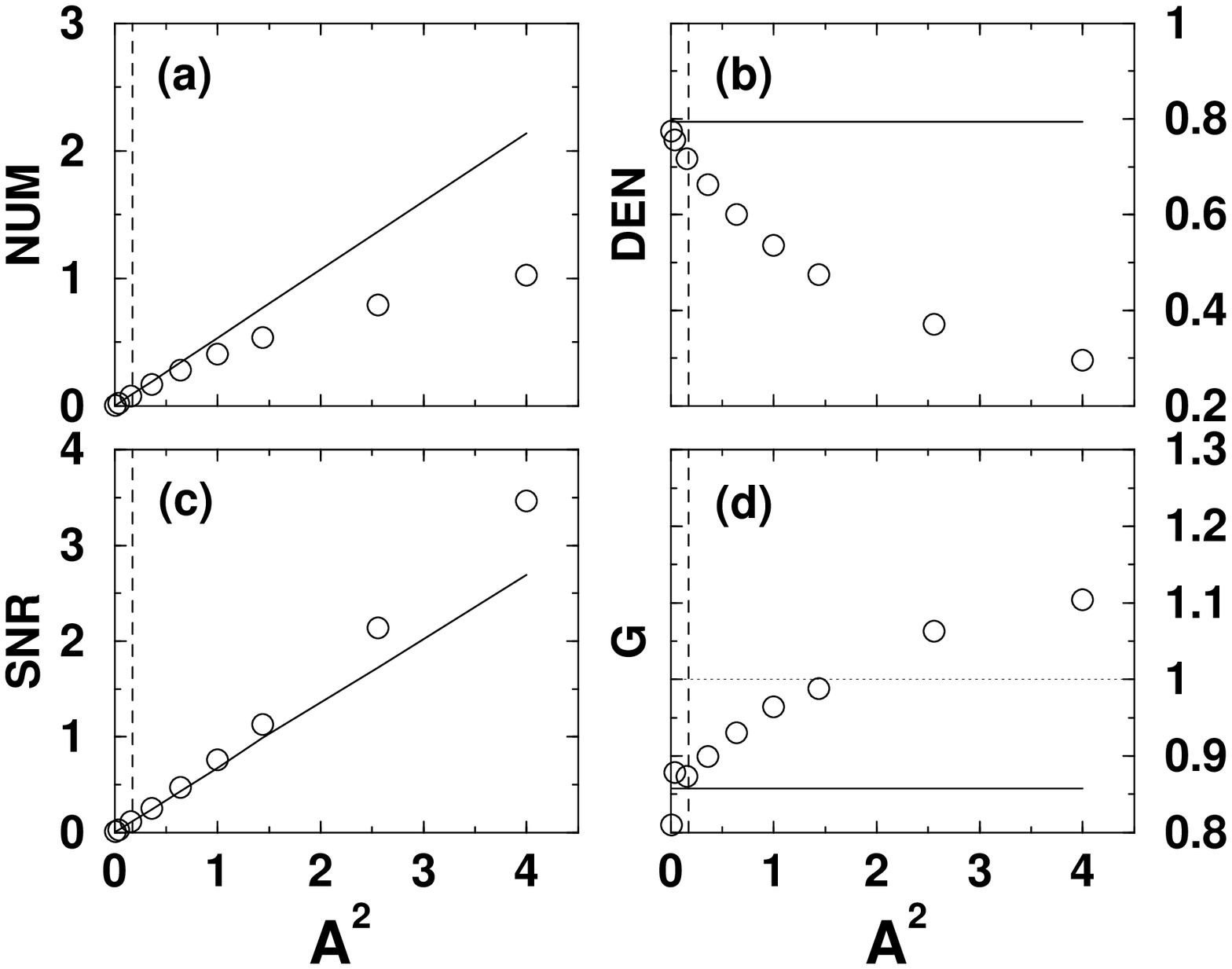}
\caption{\label{wp3} The same as in Fig.\ \ref{wp1} but now for
$\Omega=0.1$ and  $D=1.0$.}
\end{figure}

In Figs. \ref{wp2} and
\ref{wp3} we analyze the same
quantities as in Fig.\ \ref{wp1}, but now for larger noise
values, $D=0.6$ and $D=1.0$, respectively. The most important
difference with respect to the plots in Fig.\ \ref{wp1} is that
for these larger values of the noise, the gain can {\em exceed}
unity for values of the amplitude well above its threshold value.
This superthreshold feature has been corroborated already
 in Ref.~\cite{haninc00}; a gain above $1$ seemingly does not
 occur for monochromatic subthreshold driving.

\subsection{Pulsed, multichromatic  periodic driving}

\begin{figure}
\includegraphics[width=10cm]{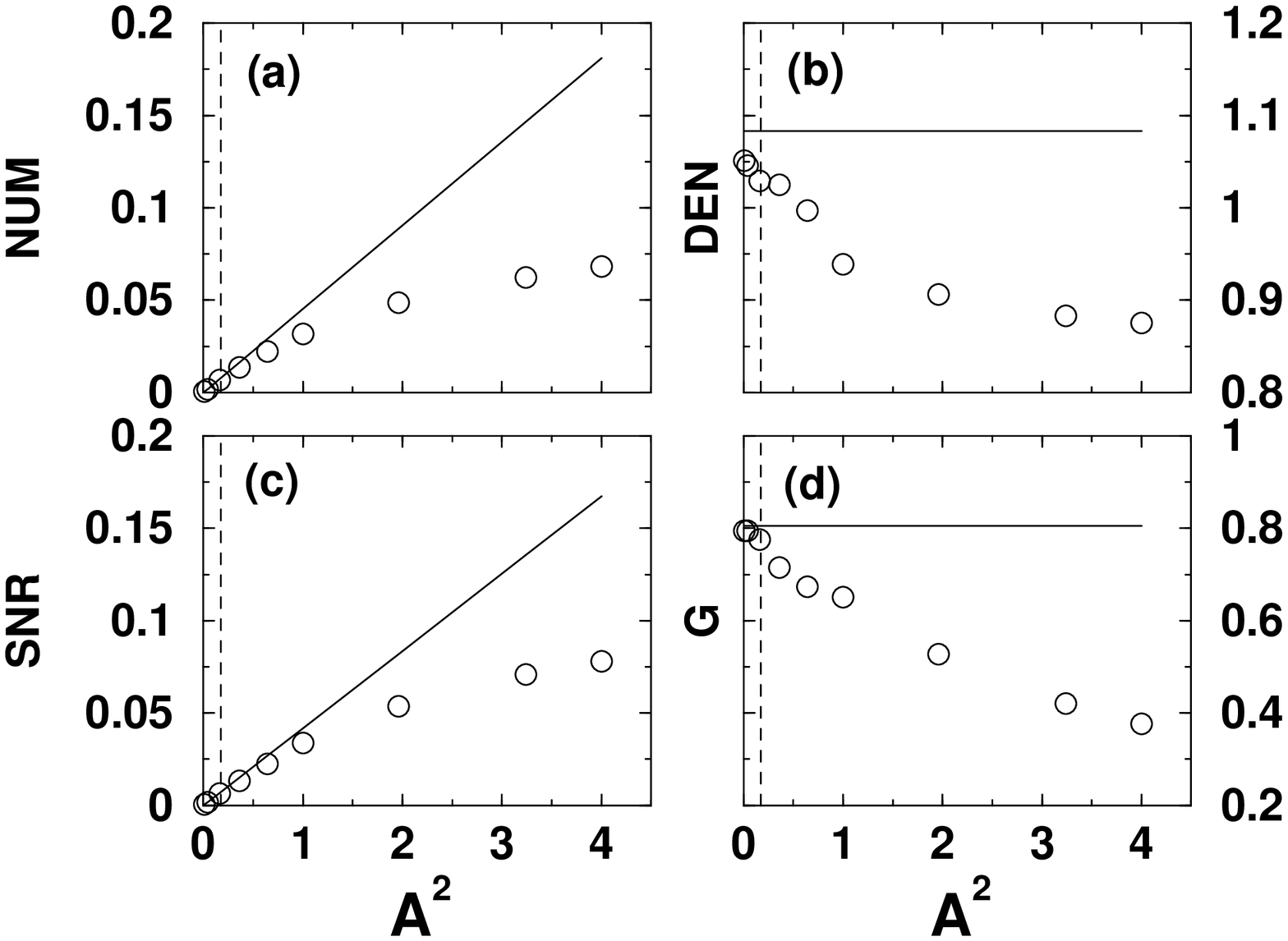}
\caption{\label{wp4} The same as in Fig.\ \ref{wp1}, for the case of a
pulsed, driving force with period $T\simeq 63$, duty cycle $2t_c/T=0.1$
cf. Eq.\ (\ref{pulse}), and a noise strength $D=0.6$.}
\end{figure}

\begin{figure}
\includegraphics[width=10cm]{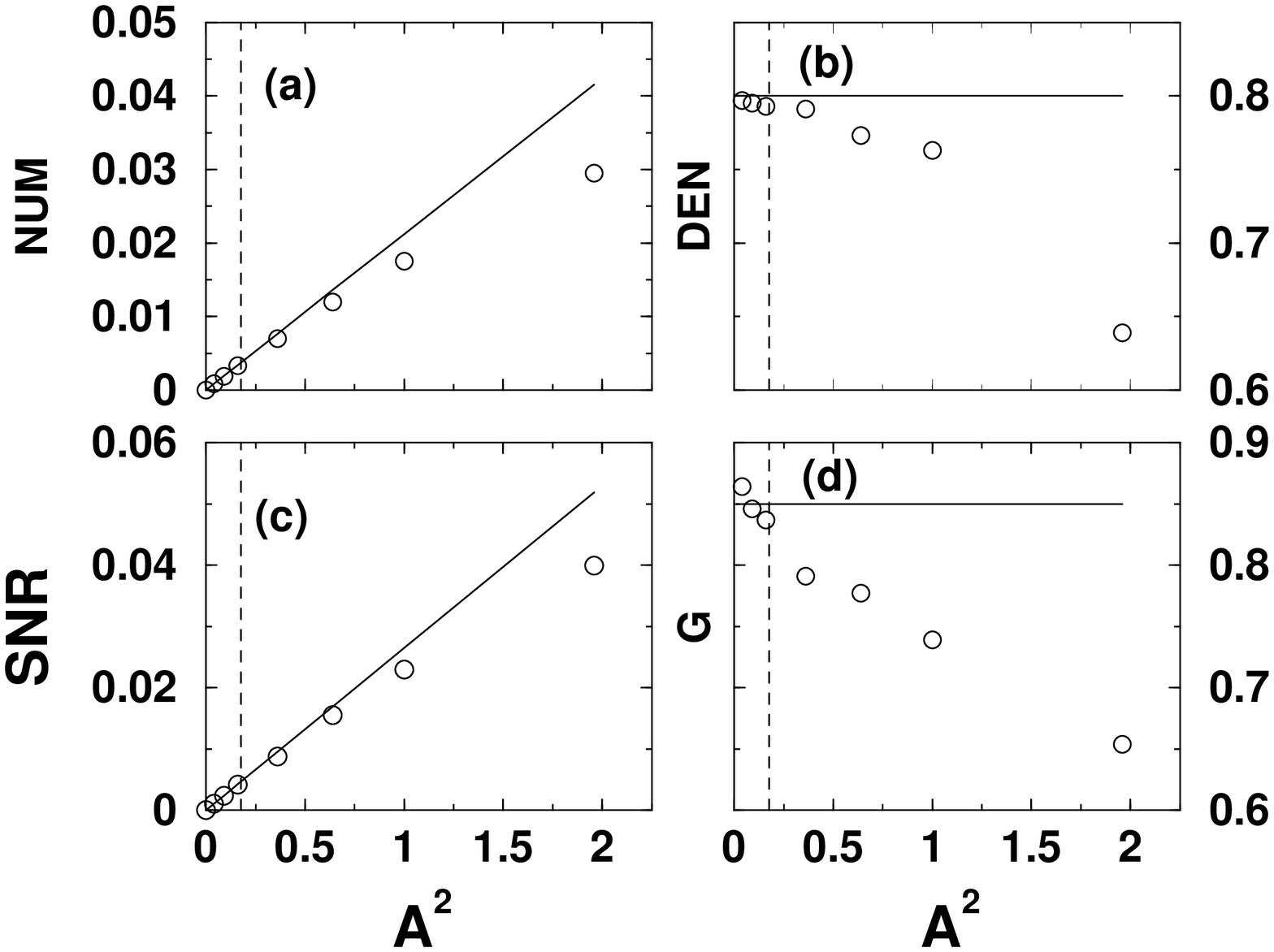}
\caption{\label{wp5}The same as in Fig.\ \ref{wp1} for a pulsed driving
force with period $T\simeq 63$, duty cycle $2t_c/T=0.1$, cf. Eq.\
(\ref{pulse}), and a noise strength $D=1.0$.}
\end{figure}

Next, we proceed to consider the case of pulsed driving forces. In
Figs.\ \ref{wp4} and \ref{wp5} we compare the dependence of
the output  on the driving amplitude as given by the LRT
approximation with the numerical precise results. The system is
forced by a multifrequency driver with a period $T=2\pi/0.1\simeq 63 $ and
a duty cycle of $10\%$.  As in the case of a single frequency
driving, the values of the different quantities obtained from the
numerics deviate significantly from the predictions of LRT as the
amplitude of the driver is increased.  Nevertheless, perhaps the
most relevant difference with respect to the monochromatic case is
that we again do {\em not} find gains larger than $ 1 $ in the
range of parameter values considered in these figures.

\subsection{The case of strong nonlinearity}

A particularly interesting situation arises in the analogue studies of
pulsed driving forces with a very {\em small} fundamental frequency: in
Refs. \cite{ginmak00,ginmak01} Gingl {\em et al.} report gains that
significantly exceed the value $1$ for a subthreshold, multifrequency
driving force of very large period $T=2\pi/0.0024\simeq 2618 $ and a
small duty cycle of $10\%$. This large gain is accompanied by a
non-monotonic behavior of the SNR with the noise strength
$D$. Therefore, this situation must correspond to a very sensible
discrepancy of the actual behavior with respect to the LRT
predictions. We have carried out detailed and careful numerics of the
Langevin equation in this extreme regime for such a driving force with a
subthreshold amplitude $A=0.35$ and a noise strength $D=0.02$. With the
parameters considered, the problem becomes computationally very
demanding indeed: this is so because of the very large period of the
driving force. Moreover, in order to obtain reliable numerical results
for the incoherent part of the correlation function a large number of
stochastic trajectories needs to be generated. Our findings are
summarized with the Table \ref{table1}.
\begin{table}
\caption{\label{table1}}
\begin{ruledtabular}
\begin{tabular}{|c||c|c|c|c|c|}
&trajectories&numerator&denominator&SNR&gain \\ \hline
numerics&1000&0.78&0.33&2.32&12.16\\ \cline{1-1}
&5000&0.78&0.35&2.26&11.84\\ 
&10000&0.78&0.47&1.67&8.77\\ 
&50000&0.78&0.48&1.65&8.62\\ \cline{1-6}
LRT&&0.00061&0.177&0.0034&0.018
\end{tabular}
\end{ruledtabular}
\end{table}

To obtain a reliable convergence of the corresponding SR quantifiers at
least up to 50000 random trajectories need to be considered. A smaller
sampling size can induce severe errors, see in Table I. The main result
is a numerically evaluated {\em gain} of $8.62$; in clear contrast, the
result predicted by LRT is the very small value of $0.018$; i.e. LRT
strikingly fails, cf. in Table I for the corresponding values of SNR and
its constituents. The SNR value of the analogue simulation in
Refs. \cite{ginmak00,ginmak01} carried out with a pulsed input signal
with the same characteristics as the one considered here, and wideband
Gaussian noise with a related strength roughly similar to ours, yields
an experimentally determined gain of ca. $19$, cf. Fig. 4 in
\cite{ginmak01}.  This value is again significantly larger than $1$ and
compares favorably with our results in Table I. Note, however, that the
sampling size of ca. $1000$ realizations used in
Refs. \cite{ginmak00,ginmak01} has been chosen substantially smaller
than the number of realizations needed to achieve good numerical convergence,
cf. Table \ref{table1}; this in turn may explain the overshoot of the
experimentally determined gain value.

\section{Conclusions}

Let us summarize  the main results of this work: (i) First,
  we have provided an analytical proof based on LRT beyond the
 commonly employed  two-mode approximation
 that the {\em gain} of a noisy,
 periodically driven nonlinear system which operates within
the regime of validity of LRT cannot exceed unity. This result
holds for arbitrary noise strength $D$ and  is independent of the
shape of the input signal. (ii) We have implemented a very
efficient algorithm due to Greenside and Helfand
\cite{hel79,grehel81} to numerically integrate the Langevin
equation. From the numerical solution, we have evaluated the time
evolution of the correlation function and its coherent and
incoherent components. (iii) We have also put forward  a
procedure, alternative to the usual one, to calculate the SNR. The
numerator and denominator of the SNR are calculated by use of only
two numerical quadratures. (iv) A detailed comparison between the
predictions of LRT and the numerical results have been carried
out. We have assessed regions of parameter values where LRT gives
an erroneous description, yet the gain,  nevertheless, is less
than unity. On the other hand, there exist regions in parameter
space where the gain indeed exceeds $1$ if driven with a
superthreshold amplitude strength; this finding is in agreement
with prior results in Ref. \cite{haninc00}. These regions are
again characterized by substantial deviations from LRT.

Moreover, as previously established by use of analogue simulations
by Gingl {\em et al.} \cite{ginmak00,ginmak01} we also find the
surprising result, {\em valid for dynamical systems}, that
SNR-gains larger than unity can indeed occur  for subthreshold (!)
polychromatic input signals: For this feature to occur one
seemingly needs, however, weak noise and a  slow periodic driving
signal with a very small duty cycle. In this context, the
necessity of a sufficiently large number of sampling trajectories
in order to obtain reliable, convergent results has also been
stressed. It is in this very regime of small frequency driving and
weak noise where the LRT description indeed  fails notably
\cite{EPLcasado,FNLcasado}.

\appendix
\section{The method of Greenside and Helfand}
The procedure proposed by Greenside and Helfand for numerically
integrating stochastic differential equations has been discussed
in detail by their authors in \cite{hel79,grehel81}.  For the sake
of completeness, we will briefly sketch in this Appendix the main
reasoning of their procedure. By analogy with deterministic
Runge-Kutta algorithms, Greenside and Helfand developed schemes to
estimate the value of the stochastic variable at time $t+h$ if its
value at time $t$ is known. This is achieved by evaluating the
right hand side of the Langevin equation at selected points within
each interval of length $h$, so that, all moments of $x(t+h)-x(t)$
are correct to order $h^k$.

As our Langevin equation contains an explicit time dependent driving
force, it is convenient to rewrite it as a two-dimensional problem with
variables $(y_1,y_2)=\vec{y}$, where $y_1=x$ and $y_2=t$. The Langevin
equation, Eq.\ (\ref{langev}) is then written in vector form as
\begin{equation}
\label{a1}
\frac {d\vec{y}}{dt}=\vec{G}(\vec{y})+\vec{\Xi}(t)
\end{equation}
where $\vec{G}= (G_1, G_2)= (-U^{'}(x)+F(t), 1)$ and $\vec{\Xi}(t)=(\xi(t),0)$.

The formal solution of Eq.\ (\ref{a1}) yields
\begin{equation}
\label{a2}
y_{\kappa}(h)=y_{\kappa}(0)+ \int_0^{h} ds\; G_{\kappa}(\vec{y}(s)) +
w_{\kappa}^{(0)}(h), \; (\kappa=1,2)
\end{equation}
with
\begin{equation}
\label{a3}
w_{\kappa}^{(0)}(h) = \int_0^{h} ds\; \Xi_{\kappa}(s)
\end{equation}
The right hand side of Eq.\ (\ref{a2}) can be expanded as
\begin{equation}
\label{a4}
y_{\kappa}(h)=y_{\kappa}(0)+ h G_{\kappa}(\vec{y}(0)) + \frac 12 h^2
\sum_{\mu}\frac {\partial G_{\kappa}(\vec{y}(0))} {\partial
y_{\mu}}G_{\mu}(\vec{y}(0))+ \ldots+ S_{\kappa}(h)
\end{equation}
The last term, $S_{\kappa}(h)$, represents the stochastic part. It is a
series in $h^{1/2}$ with the order of the terms determined in
probability.

By analogy with the Runge-Kutta procedures for deterministic
differential equations, Greenside and Helfand propose an $l$-stage
algorithm to write the solution of Eq.\ (\ref{a1}) as
\begin{equation}
\label{a5}
y_{\kappa}(h)=y_{\kappa}(0)+h(A_1 g_{1\kappa}+\ldots+A_lg_{l\kappa}) +
h^{\frac12} \Xi_{\kappa}^{\frac12} Y_{0\kappa}
\end{equation}
with
\begin{eqnarray}
\label{a6}
g_{1\kappa}&=&G_{\kappa}\left (\left \{ y_{\mu}(0)+h^{\frac12}
\Xi_{\mu}^{\frac12} Y_{1\mu}\right \} \right )\nonumber\\
g_{2\kappa}&=&G_{\kappa}\left (\left \{ y_{\mu}(0)+h\beta_{21}g_{1\mu}+h^{\frac12}
\Xi_{\mu}^{\frac12} Y_{2\mu}\right \} \right )\nonumber\\
\vdots \nonumber \\
g_{l\kappa}&=&G_{\kappa}\left (\left \{
y_{\mu}(0)+h\beta_{l1}g_{1\mu}+\ldots+h\beta_{l,l-1}g_{l-1,\mu}+ h^{\frac12}
\Xi_{\mu}^{\frac12} Y_{l\mu}\right \} \right)
\end{eqnarray}
Here, $(\{ y_{\mu} \})$ is the set $(x,t)$. The $Y_{l\mu}$ are Gaussian
stochastic variables with zero average which are numerically generated
by writing
\begin{equation}
\label{a7}
Y_{i\kappa}=\sum_{j=1}^m \lambda_{ij} Z_{j \kappa}
\end{equation}
where $Z_{j \kappa}$ are $m$ independent Gaussian random variables
of zero average and unit variance. The parameters $A_i$,
$\beta_{ij}$ and $\lambda_{ij}$ appearing in Eqs.\
(\ref{a5}-\ref{a7}) are independent of the component index
$\kappa$. They are obtained by expanding Eq.\ (\ref{a5}) to the
desired order $h^k$. This expansion gives rise to a deterministic
and a stochastic part, $\tilde{S}_{\kappa}$. Equating the
coefficients of this expansion with those of the deterministic
part in Eq.\ (\ref{a4}) leads to a set of equations for the
parameters  $A_i$, $\beta_{ij}$ and  $\lambda_{ij}$. Further
equations are obtained by equating the moments of $\langle
\tilde{S}_{\kappa}^n \rangle$ with those of the stochastic part in
the expansion in Eq.\, (\ref{a4}) $\langle S_{\kappa}^n \rangle$.

A procedure correct to order $h^k$ in the step size $h$, involving
$l$-stages and $m$ Gaussian independent variables is termed a
$k_Ol_Sm_G$ algorithm. In this paper we have integrated the Langevin
equation using a $3_O4_S2_G$ algorithm with the values for $A_i$,
$\beta_{ij}$ and $\lambda_{ij}$ given in Table \ref{table2} taken from
\cite{grehel81}. With this choice of parameters, the deterministic part
is of order $h^4$, as in the fourth order Runge-Kutta procedure for
ordinary differential equations.
\begin{table}
\caption{Parameter values given by Greenside and Helfand \cite{grehel81}
for their $3_O4_S2_G$ algorithm\label{table2}}
\begin{ruledtabular}
\begin{tabular}{|c|c|c|c|}
A$_1$&0.0&A$_2$&0.644468\\
A$_3$&0.194450&A$_4$&0.161082\\
$\beta_{21}$&0.516719&$\beta_{31}$&-0.397300\\
$\beta_{32}$&0.427690&$\beta_{41}$&-1.587731\\
$\beta_{42}$&1.417263&$\beta_{43}$&1.170469\\
$\lambda_{01}$&1.0&$\lambda_{02}$&0.0\\
$\lambda_{11}$&0.0&$\lambda_{12}$&0.271608\\
$\lambda_{21}$&0.516719&$\lambda_{22}$&0.499720\\
$\lambda_{31}$&0.030390&$\lambda_{32}$&-0.171658\\
$\lambda_{41}$&1.0&$\lambda_{42}$&0.0\\
\end{tabular}
\end{ruledtabular}
\end{table}
\begin{acknowledgments}
We acknowledge the support of the Direcci\'on General de
Ense\~nanza Superior of Spain (BFM2002-03822), the Junta de
Andaluc\'{\i}a, the DAAD program "Acciones Integradas" (P.H., M.
M.) and the Sonderforschungsbereich 486 of the Deutsche
Forschungsgemeinschaft, project A10.
\end{acknowledgments}

\newpage

\end{document}